\documentclass[%
 reprint,
 superscriptaddress,
 amsmath,amssymb,
 aps,
 prl,
 floatfix
]{revtex4-2}

\usepackage{graphicx}
\usepackage{dcolumn}
\usepackage{bm}
\usepackage[colorlinks=true,allcolors=blue]{hyperref}
\usepackage{physics}
\usepackage{xspace}
\usepackage{xcolor}
\usepackage[T1]{fontenc}
\usepackage{newtxtext,newtxmath}
\usepackage{color}
\usepackage{url}
\usepackage{comment}
\usepackage{float}
\usepackage{soul}
\usepackage{ulem}
\begin{document}

\title{Topological Invariants in Nonlinear Thouless Pumping of Solitons}
\makeatletter\let\thetitle\@title\makeatother

\author{Fei-Fei~{Wu}}

\author{Xian-Da~{Zuo}}
\author{Qing-Qing~{Zhu}}
\author{Tao~Yuan}
\author{Yi-Yi Mao}
\affiliation{Hefei National Research Center for Physical Sciences at the Microscale and School of Physical Sciences, University of Science and Technology of China, Hefei 230026, China}
\affiliation{Shanghai Research Center for Quantum Sciences and CAS Center for Excellence in Quantum Information and Quantum Physics, University of Science and Technology of China, Shanghai 201315, China}

\author{Chao~Zeng}
\affiliation{Hefei National Research Center for Physical Sciences at the Microscale and School of Physical Sciences, University of Science and Technology of China, Hefei 230026, China}
\affiliation{Shanghai Research Center for Quantum Sciences and CAS Center for Excellence in Quantum Information and Quantum Physics, University of Science and Technology of China, Shanghai 201315, China}
\affiliation{Hefei National Laboratory, University of Science and Technology of China, Hefei 230088, China}

\author{Yi~{Jiang}}
\affiliation{Hefei National Research Center for Physical Sciences at the Microscale and School of Physical Sciences, University of Science and Technology of China, Hefei 230026, China}
\affiliation{Hefei National Laboratory, University of Science and Technology of China, Hefei 230088, China}
\author{Yu-Ao~{Chen}}
\affiliation{Hefei National Research Center for Physical Sciences at the Microscale and School of Physical Sciences, University of Science and Technology of China, Hefei 230026, China}
\affiliation{Shanghai Research Center for Quantum Sciences and CAS Center for Excellence in Quantum Information and Quantum Physics, University of Science and Technology of China, Shanghai 201315, China}
\affiliation{Hefei National Laboratory, University of Science and Technology of China, Hefei 230088, China}
\affiliation{New Cornerstone Science Laboratory, School of Emergent Technology, University of Science and Technology of China, Hefei 230026, China}
\author{Jian-Wei~{Pan}}
\author{Wei~{Zheng}}
\email[]{zw8796@ustc.edu.cn}
\author{Han-Ning~{Dai}}
\email[]{daihan@ustc.edu.cn}

\affiliation{Hefei National Research Center for Physical Sciences at the Microscale and School of Physical Sciences, University of Science and Technology of China, Hefei 230026, China}
\affiliation{Shanghai Research Center for Quantum Sciences and CAS Center for Excellence in Quantum Information and Quantum Physics, University of Science and Technology of China, Shanghai 201315, China}
\affiliation{Hefei National Laboratory, University of Science and Technology of China, Hefei 230088, China}

\date{\today}


\begin{abstract}
\noindent 

Recent explorations of quantized solitons transport in optical waveguides have thrust nonlinear topological pumping into the spotlight. In this work, we introduce a unified topological invariant applicable across both weakly and strongly nonlinear regimes. In the weak nonlinearity regime, where the nonlinear bands are well-separated, the invariant reduces to the Abelian Chern number of the occupied nonlinear band. Consequently, the pumped charge is quantized to an integer value. As the nonlinearity increases, the nonlinear bands start to intertwine, leading to a situation where the invariant is expressed as the non-Abelian Chern number divided by the number of interacting bands. This could result in a fractional quantization of the pumped charge. Our unified topological invariant approach not only advances the understanding of the soliton dynamics, but also provides implications for the future design of nonlinear topological systems.

\end{abstract}

\maketitle


Topological Thouless pumping, leveraging adiabatic and periodic modulation of one-dimensional lattices, offers a unique method for charge transport in the absence of directed external fields~\cite{Thouless1983, Niu1984, Niu1990, Citro2023}. Such a pumping process can be mapped to the two-dimensional topological Chern insulator. Therefore the pumped charge is quantized according to the topological invariant of the corresponding two-dimensional energy bands. Recent experiments have implemented Thouless pumping in various platforms, including ultracold atoms in optical lattices~\cite{Lohse2016, Nakajima2016, Lu2016}, photonic~\cite{Kraus2012, Zilberberg2018},  acoustic~\cite{Cheng2020}, and plasmonic systems~\cite{Fedorova2020}, as well as superconducting processors~\cite{Deng2024}. Besides, significant efforts have explored the Thouless pumping in non-Abelian case~\cite{Brosco2021,Wu2019,You2022} and strongly interacting regimes~\cite{Tangpanitanon2016,Leechaohong2020,Leechaohong2023,Walter2023,Leechaohong2023b,Viebahn2023,Liu2023}.

Recent studies of topological materials have expanded into nonlinear systems~\cite{Smirnova2020,Zhou2022}. Nonlinearity arises naturally as one investigating light propagation in waveguide lattices  made of nonlinear, or the dynamics of Bose-Einstein condensates in optical lattices. Nonlinearity can also be introduced to electronic LC circuits via varactor diodes. In these materials, the interplay between nonlinearity and topology can support self-localized solitons on edge or in the bulk~\cite{Mukherjee2020,Mukherjee2021,Lumer2013,Leykam2016,Marzuola2019}. Self-induced topological transitions have been theoretically predicted in nonlinear phononic crystals~\cite{Chaunsali2019}. Meanwhile, the classification of nonlinear topological matters has been explored via numerical K-theory~\cite{Wong2023}. 

Thouless pumping, as a topological phenomenon in the space-time domain, has also been extended to nonlinear regimes~\cite{Jurgensen2021,Jurgensen2022,Jurgensen2023,Fu2022,Tuloup2023,Fu2022a,Mostaan2022,Yuan2023,Leechaohong2025,Ravets2025,Cao2024,Zhang2009,Tao2025,Cao2025}. For example, a quantized pumping of a single soliton excitation has been demonstrated in a one-dimensional Aubry-André-Harper (AAH) model incorporating Kerr nonlinearity~\cite{Jurgensen2021}.
More interestingly, extending the cell size at moderate nonlinearity strengths leads to the emergence of fractional pumping of the soliton~\cite{Jurgensen2023}.
Several theoretical frameworks have been proposed to explain the quantized pumping of solitons~\cite{Fu2022, Jurgensen2022, Mostaan2022, Tuloup2023}, either by expanding the solitons on the basis of linear bands, or through the nonlinear bands in the weak nonlinear regime~\cite{Sone2024}. However, strong nonlinearity will induce Kerr loops in the energy bands~\cite{Wu2000}. These loops subsequently expand and intertwine with each other, forming "{\it braiding}" bands. In this situation, the picture of pumping based on linear or weak nonlinear bands breaks down.

\begin{figure}[htb]
    
	\includegraphics{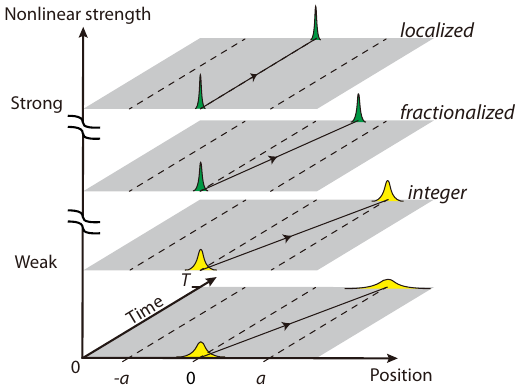}
	\caption{\label{Fig1} 
    Schematic of the nonlinear Chern number and the corresponding topological phase in 1+1D system with Kerr-like nonlinear.
    The system is of translation symmetries and driven by some periodical parameters.
    In a linear system, the pumping charge of the Thouless pumping is the first Chern number.
    For nonlinear situations, the nonlinear Chern number is determined by nonlinear eigenvectors.
    In a weak regime, the exciton is pumped out with the charge related to the Chern number of the nonlinear band.
    Beyond the weak regime, the exciton is localized agreed with the zero nonlinear Chern number, when the system digs into a very strong regime.
    There also could emerge fractional quantized pump charge in a moderate nonlinearity. 
}
\end{figure}

In this work, we provide a topological invariant for nonlinear Thouless pumping of solitons. Our invariant is valid even in the strong nonlinear regime, where nonlinear bands "braiding" to each others. Therefore, our invariant can describe both integer pumping in weak nonlinear regime and fractional pumping in strong nonlinear regime. We found that the displacement of a soliton after one nonlinear pumping circle is given by
\begin{equation}\label{eq:one}
    D = \frac{{C_{{\mathrm{NL}}} }}{N},
\end{equation}
where $C_{\mathrm{NL}}$, unlike other works, is the non-Abelian Chern number of the lowest {\it braiding} (degenerate) nonlinear bands. $N$ is the number of these {\it braiding} bands due to the nonlinearity. When the system is linear and the lowest bands are well separated from other bands, this formula is reduced into traditional Thouless pumping, $D=C_{\mathrm{Linear}}$. The displacement is given by the Abelian Chern number of the lowest linear band. In the weak nonlinear regime, $C_{\mathrm{Linear}}$ should be replaced by the Chern number of nonlinear bands, thus giving the integer nonlinear pumping. As increasing the nonlinearity, Kerr loops start to emerge in the nonlinear bands. In this situation, the topology of the nonlinear bands breaks down, and the pumping of a soliton is not quantized. When the nonlinearity becomes larger, Kerr loops expand, and form {\it braiding} bands. These degenerate bands intertwine with each other, and can not be well separated. In this case, the displacement is given by the non-Abelian Chern number of these bands divided by the number of bands involved. That describes the fractional nonlinear pumping of solitons, see Fig.~\ref{Fig1}.


\textit{Chern number of nonlinear bands}. As shown in Fig.~\ref{Fig1}, considering a 1D system with periodically driven parameters $\mathcal{R}(t,x)$, the nonlinear term is given by $g|\Psi(t,x)|^2$, i.e., Kerr-type nonlinearity.
The time-dependent Hamiltonian is written as,
\begin{equation} \label{eq:two}
    \begin{aligned}
    H(t,\Psi(t,x)) \!=\! H_{\mathrm{lin}}(t,x) - diag (g|\Psi(t,x)|^2),
    \end{aligned}
\end{equation}
where $H_{\mathrm{lin}}$ is the linear Hamiltonian and $\Psi$ is the nonlinear eigenvector.
$g$ denotes the nonlinear strength and $diag(g|\Psi(t,x)|^2)$ is a diagonal matrix.
The dynamic evolution of Eq.~(\ref{eq:two}) is described by the nonlinear Schrödinger equation, $i \hbar \partial_t\Psi(t,x)\!=\!H(t,\Psi(t,x)) \ \Psi(t,x)$.
The pumping charge is defined by the total displacement of the center of mass (COM) over a single period, $\mathbf{D} \!=\![\sum x|\Psi(t=\mathrm{T},x)|^2  - \sum x|\Psi(t=0,x)|^2]/a$, where $a$ is the size of the unit cell, and $\mathrm{T}$ is the period.
Above a certain $g$, solitons will form with special initial states, i.e., ground states of the nonlinear Schrödinger equation.
As $g$ increases, pumping charges of solitons transition from integer to localization, with fractional pumping potentially occurring between them.
Hereafter, we discuss the pumping for the initial states occupying only a single eigenvector to explore the relation between the nonlinear Chern number and pumping charge.

\begin{figure*}[t]
	\includegraphics{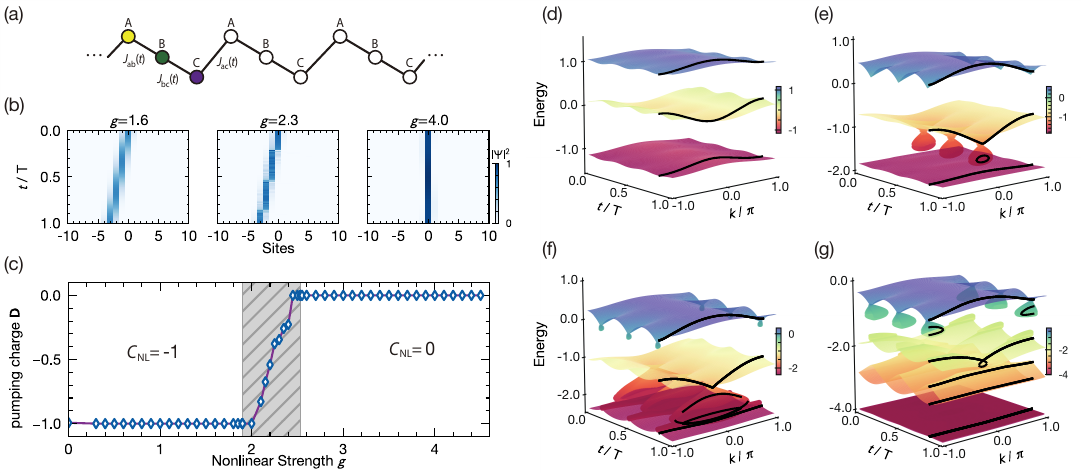}
	\caption{\label{Fig2} 
    (a), The diagram of the linear AAH model with three sites per unit cell. $J_{ab}(t),~J_{bc}(t)$, and $J_{ac}(t)$ are the periodical modulated tunneling strength between corresponding sites.
    (b), The simulations of Thouless pumping with different nonlinear strengths.
    (c), The phase diagram and the pumping charge of the 3-site nonlinear AAH model.
    The horizontal axis is the nonlinear strength.
    The diamonds are the pumping charge $\mathbf{D}$ over one cycle. 
    The gray shaded area is the ill-defined region of the topological Chern number.
    (d)-(g), Energy bands of 3-site AAH model for $g\!=\!0.0$, $g\!=\!1.6$, $g\!=\!2.3$ and $g\!=\!4.0$ in 2D BZ, respectively.
    The black lines in (d)-(g) are the eigenvalues for $t\! =\! 1.725\pi$.
    (d), In the linear model, there are three bands with Chern number \{-1, 2, -1\}.
    (e), For $g\!=\!1.6$, loop structures emerge but the lowest band stay isolated.
    (f), For $g\!=\!2.3$, loop structures expand so that they touch the lowest band.
    (g), For $g\!=\!4.0$, in a large energy scale, more loops are generated, and the loops in (b) have expanded to all 2D BZ.
    The tunneling parameters are $J\! =\! 1.2$, $K\! = \!1.0$.
}
\end{figure*}

Under the adiabatic approximation, $\Psi(t,x)$ is the unitary gauge transformation of instantaneous nonlinear eigenvectors of $H(t,\Psi(t,x))$.
In addition, considering the spatial translation symmetries, we can describe the time-dependent wave function in terms of instantaneous Bloch states in momentum space, $\Psi(x)=\Phi(k) e^{i k x}$.
The instant nonlinear eigenvectors $\Phi(k)$ and eigenvalues $E(k)$ are solved by the modified Newton method (see Supplemental Material~\cite{Supplementary}).
According to the nonlinearity, loop structures emerge in energy bands within certain regions of the 2D first Brillouin Zone(BZ), suggesting that the number of nonlinear eigenvectors may exceed the dimension of the Hamiltonian.
This indicates that the nonlinear eigenvectors are not orthogonal.
These loops expand as the nonlinear strength increases, eventually filling the entire first BZ.
Degeneracy may occur for large $g$, leading to a situation beyond Abelian conditions thus the standard Abelian Chern number no longer works.
The degenerate subspace is spanned by $N$ degenerate eigenvectors $[\Phi_1,\Phi_2,\Phi_3,...,\Phi_N]$.
Despite the non-orthogonality, they still preserve $\mathrm{U}(N)$ gauge due to the normalization of eigenvectors, $\sum_j|\phi_j(k)|^2\! =\! 1$.
We define the nonlinear Chern number using the non-Abelian Berry connection,
\begin{equation}\label{eq:three}
    \begin{aligned}
    C_{\mathrm{NL}}\! =\! \frac{1}{2 \pi i} \int_0^T d t \int_{-\pi}^\pi d k \mathrm{Tr} F(\mathbf{k}),
    \end{aligned}
\end{equation}
where $F(\mathbf{k})=\partial_t \mathbf{A}_k-\partial_k \mathbf{A}_t-i\left[\mathbf{A}_t, \mathbf{A}_k\right]$ is the non-Abelian Berry curvature.
$\mathbf{A}_k$ is the one-form Wilczek–Zee or Mead-Berry connection~\cite{Wilczek1984} with matrix elements $[\mathbf{A}_k]_{ab}\!=\!i\left\langle\Phi_a(\mathbf{k}) \mid\partial_k \Phi_b\left(\mathbf{k}\right)\right\rangle$ and the integration is over 1+1D first BZ of $\mathbf{k} \!= \!\{k,t\}$.
The $N\times N$ matrix $\mathbf{A}_{k}$ reduces to the Abelian Berry connection $\mathbf{A}(\mathbf{k})\!=\!i\left\langle\Phi(\mathbf{k}) \mid \nabla_{\mathbf{k}} \Phi(\mathbf{k})\right\rangle$ for $N=1$, corresponding to the non-degenerate situation. Here, we emphasize there is no limitation on the nonlinear strength in our method, and Eq.~(\ref{eq:three}) also applies for braiding bands in the regime with large nonlinearity.

\textit{Nonlinear AAH model.} In the following discussion, we explore the topological phase transition induced by the nonlinearity and the corresponding nonlinear Chern number in the 1+1D AAH model.
According to the Bloch theorem, the instantaneous nonlinear Hamiltonian for Fig~\ref{Fig2}(a) in momentum space is,
\begin{equation}\label{eq:four}
    \begin{aligned}
    H(k, t)\!=\!\left(\begin{array}{ccc}-g\left|\phi_1(k,t)\right|^2 & J_{a b}(t) & J_{c a}(t) e^{-i k} \\ J_{a b}(t) & -g\left|\phi_2(k,t)\right|^2 & J_{b c}(t) \\ J_{c a}(t) e^{i k} & J_{b c}(t) & -g\left|\phi_3(k,t)\right|^2\end{array}\right),
    \end{aligned}
\end{equation}
where $\Phi(k,t)\! =\! [\phi_1(k,t),\phi_2(k,t),\phi_3(k,t)]^T$ represents the eigenvector at time $t$ and momentum $k$.
The periodical tunneling strength between the nearest neighbor sites $A$ and $B$ follows a cosine modulation, $J_{ab}(t) \!=\! \left[J+K \cos \left(t+2 \pi/6\right)\right]/(J+K)$ where $J$ and $K$ are tunneling parameters. $J_{bc}(t)$ and $J_{ac}(t)$ are delayed by $2\pi/3$ and $4\pi/3$, respectively.
We simulate the dynamic evolution of the nonlinear Hamiltonian in real space.
The evolution for $g=1.5$, $g=2.3$ and $g=3.0$ is shown in Fig~\ref{Fig2}(b).
The pumping charges for $g=0.0$ to $4.5$ are shown in Fig~\ref{Fig2}(c).
For weak $g$, $\mathbf{D}=-1$; for strong $g$, localized solitons are formed, thus $\mathbf{D}=0$.
At medium $g$, no stable solitons exist, and the pumping charge is somewhat random and not quantized.

The energy bands calculated through the Newton downhill method are shown in Fig.~\ref{Fig2}(d-g) for $g=0.0$, 1.6, 2.3, and 4.0.
As the nonlinear strength $g$ increases, richer and more complex energy structures appear. 
For the linear case, $g\!=\!0$, there are three bands as shown in Fig.~\ref{Fig2}(d), with Chern numbers of $\{-1,2,-1\}$.
The pumping charge for each eigenvector is equal to its Chern number in the linear model of Thouless pumping. 
In the weakly nonlinear regime, the lowest band remains isolated, although the Kerr-loop structures emerge in certain regions of the 2D BZ, shown in Fig.~\ref{Fig2}(e).
The nonlinear Chern number of the ground state is $C_{\mathrm{NL}}\!=\!-1$.
With further increases in strength $g$, the following phenomena occur: 
1) more loops emerge;
2) loop structures gradually expand in the 2D BZ;
3) the energy gap between the ground and the loop bands vanishes.
Above a critical threshold, called the strongly nonlinear regime, the lowest loops fill the entire 2D BZ, where the lowest three bands exhibit braiding behavior.
As shown in Fig.~\ref{Fig2}(g) for $g = 4.0$, the two lowest bands become degenerate at specific times $t\!=\!(2\pi/3, 5\pi/3)$, while the second and third lowest bands degenerate at $t\!=\!(0,\pi)$ and the lowest and third lowest bands degenerate at $t\!=\!(\pi/3,4\pi/3)$. 
Notably, the three lowest bands remain isolated from the other higher bands throughout the 2D BZ.
Thus, these bands construct a three-dimensional degenerate subspace, $\mathit{\Phi}\! = \![\Phi_0,\Phi_1,\Phi_2]^T$.
The nonlinear Chern number $C_{\mathrm{NL}}\!=\!0$ is calculated by the Wilczek–Zee Berry connection~\cite{Fukui2005}.
\begin{figure}[t]
	\includegraphics{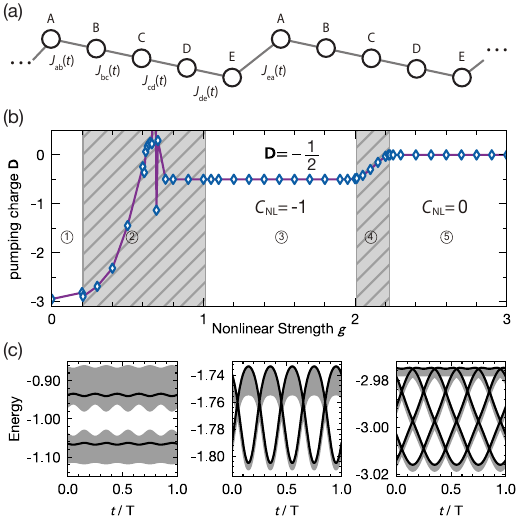}
	\caption{\label{Fig3} (a), Schematic of the unit cell with five sites.
    (b), The phase diagram and pumping charge for the nonlinear five-site AAH model.
    There are three topological phases noted by 1, 3, and 5.
    The gray-shaded regions 2 and 4 are the ill-defined areas of the topological Chern number.
    The blue diamonds are the pumping charges.
    (c), Some low energy bands for the nonlinear five-site AAH model, corresponding to the three well-defined regions in (b).
    The gray shadows are the projection of bands over $k$.
    The black lines are the energy bands at $k\!=\!0.5\pi$.
    From left to right is of $g\! =\! 0.0$, $g\! =\! 1.6$ and $g\! =\!  3.0$.
    For $g\! =\! 1.6$, the lowest two bands are shown. For $g\! =\! 3.0$, the lowest five bands are braiding.
    The tunneling parameters are $J\! =\! 1.2$ and $K\! =\! 1.0$ for all figures above.}
\end{figure}

Between the weak and strong regimes, there is a region of medium nonlinear strength where the energy bands exhibit more intricate structures, as shown in Fig.~\ref{Fig2}(f). 
In this regime, the lowest band and loop bands become degenerate in certain regions. 
However, these loops connect other excited states as well, and they are incompletely filled in the 2D BZ, leading to an ill-defined Chern number.
The degenerate points of filled bands and non-filled bands induce the non-adiabatic evolution and non-stable solitons~\cite{Tuloup2023}.

Furthermore, we explore the relation between the nonlinear Chern number and the pumping charges of solitons.
In the nonlinear system, the change of pumping charges indicates the phase transition.
Compared to a linear system, the pumping charge becomes more intricate with braiding nonlinear bands.
The pumping charge $\mathbf{D}$ is the average nonlinear Chern number over the dimension of degenerate subspace $N$, see Eq.~(\ref{eq:one}),
which is like the generalized Thouless-Kohmoto-Nightingale-den Nijs (TKNN) formula~\cite{Thouless1982}, describing the Hall conductance for quantum Hall effect states.
In Fig.~\ref{Fig2}(c), we find the nonlinear Chern number agrees well with the pumping charges, indicating the definition of the nonlinear topological invariant, Eq.~(\ref{eq:three}), is valid.

To illustrate the necessity of the denominator $N$ in Eq.~(\ref{eq:one}), we extend the cell size to five sites, leading to the emergence of more complicated energy bands that can induce quantized fractional pumping~\cite{Jurgensen2023}, as shown in Fig.~\ref{Fig3}.
The tunneling strength follows $\mathrm{cos}$ modulation, $J_{j,j+1}(t) \!=\! \left[J+K \cos \left(4 \pi j/5+t-3 \pi/10\right)\right]/(J+K)$.
The linear Hamiltonian in this five-site model exhibits Chern numbers of $\{-3,2,2,2,-3\}$ for the five bands.
As the nonlinear strength increases, the two lowest nonlinear bands begin to intertwine, as shown in Fig.~\ref{Fig3}(c) of $g=1.6$. 
In the dynamical simulation, two stable solitons move by $-1/2$ of a unit cell per pumping cycle.
The nonlinear Chern number of these two braiding nonlinear bands is $C_{\mathrm{NL}} \!=\! -1$.
The pumping charge can be determined as $\mathbf{D}\! =\! C_{\mathrm{NL}}/2\! =\! -1/2$, which is in excellent agreement with the results obtained from dynamical pumping simulations.
Further increasing the nonlinear strength leads to the intertwining of the five lowest bands (Fig.~\ref{Fig3}(c) of $g=3.0$). 
The nonlinear Chern number of the five-dimensional degenerate space is $C_{\mathrm{NL}}\! =\! 0$, matching the $0$ pumping charge.
Further increasing the cell size to seven (see Supplemental Material~\cite{Supplementary}), we find the nonlinear Chern number is $C_{\mathrm{NL}}\! =\! -1$ with three degenerate bands for $g\! =\! 1.5$. 
The pumping charge is $\mathbf{D}\! =\! C_{\mathrm{NL}}/3 = -1/3$, which agrees well with dynamical simulations.

\begin{figure}[t]
	\includegraphics{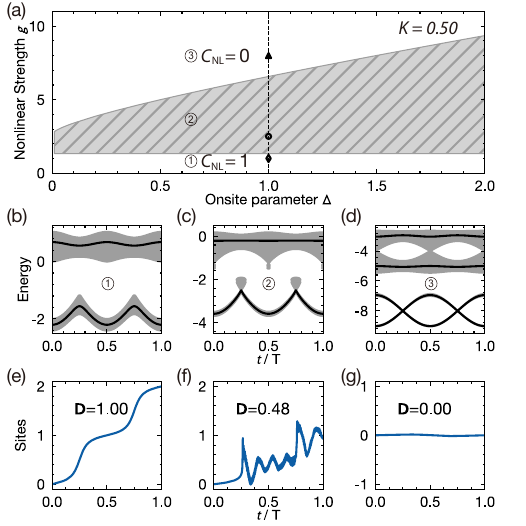}
	\caption{\label{Fig4}(a), The phase diagram for the RM model of $K\!=\!0.5$.
    The tunneling parameters are $J = 1.0$.
    The phase 1 and 3 are the well-defined topological phases.
    The gray-shaded phase 2 is the region where the topological Chern number is ill-defined.
    Energy structures (b)-(d) and simulations of the Thouless pumping (e)-(g) for the three points in (a).
    From left to right are of the point in $C_{\mathrm{NL}}=1$ (Diamond), ill-defined (Circle) and $C_{\mathrm{NL}}=0$ (Triangle) phase region in (a).
    The black lines in (b)-(d) are the energy bands at $k\!=\!0.5\pi$.
}
\end{figure}

\textit{Nonlinear Rice-Mele model.} In this section, we apply our method to a system with linear interaction.
We consider the Thouless pumping of the 1D nonlinear Rice-Mele (RM) model with staggered onsite potential.
The momentum space Hamiltonian is given by,
\begin{equation}\label{eq:six}
H(k, \Phi(k,t),t)\!=\!\left(\begin{array}{cc}V(t)-g\left|\phi_a(k,t)\right|^2 & J_1(t)+J_2(t) e^{-i k} \\ J_1(t)+J_2(t) e^{i k} & -V(t)-g\left|\phi_b(k,t)\right|^2\end{array}\right),
\end{equation}
where $J_1(t) \!=\! (J+K \mathrm{sin}(t/\mathrm{T}))/(J+K)$ is the intra-cell tunneling, and $J_2(t) \!=\! (J-K \mathrm{sin}(t/\mathrm{T}))/(J+K)$ is the inter-cell tunneling.
$V(t)\!=\! -\Delta\mathrm{cos}(t/\mathrm{T})$ represents the onsite potential.
We calculate the three-dimensional phase diagram of modulation parameters $\Delta$, $K$, and nonlinear strength $g$(see Supplemental Material~\cite{Supplementary}) and find there are two topological well-defined phases. As shown in Fig.~\ref{Fig4}(a) is the phase diagram for $K=0.5$.
One phase has nonlinear Chern number $C_{\mathrm{NL}}\!=\! 1$ (weak region 1), and the other has $C_{\mathrm{NL}} \!=\! 0$ (strong region 3), while between them is an ill-defined region 2.
The nonlinear eigenstates and eigenvalues are solved by the exact solutions.
We show the band structures and the dynamical simulations for the points in Fig.~\ref{Fig4}(a) of $g\!=\!1.0$,  $g\!=\!2.5$ and  $g\!=\!8.0$ with the same onsite parameter $\Delta\!=\!1.0$ and tunneling parameters $J\!=\!1.0$, $K\!=\!0.5$.
In the linear case, there are two separated eigenstates, which also persist for weakly nonlinear strength, Fig.~\ref{Fig4}(b).
The pumping charge in the weak region is $\mathbf{D}\!=\!C_{\mathrm{NL}}\!=\!1.0$ (Fig.~\ref{Fig4}(e)).
For strongly nonlinear strength, at most four eigenstates exist, and the lowest two bands cross at certain points in the 2D BZ, as shown in Fig.~\ref{Fig4}(d).
The states are localized $\mathbf{D}\!=\!C_{\mathrm{NL}}/2\!=\!0.0$ (Fig.~\ref{Fig4}(g)).
In the ill-defined region, ground states braid with loop structures (Fig.~\ref{Fig4}(c)), which leads to the non-adiabatic pump process (Fig.~\ref{Fig4}(f)). 
The topological phase boundary of $C_{\mathrm{NL}}\! =\! 1$ which is almost flat over $\Delta$, is related to the modulation parameters $J$ and $K$.
Changing the choice of parameters will alter both phase boundaries and can narrow the region of the ill-defined area.



\textit{Summary.} In conclusion, based on nonlinear band structures, we provide a unified topological invariant for both integer and fractional nonlinear Thouless Pumping of solitons. This invariant can be applied in a weakly nonlinear regime, where the nonlinear band is well separated, and gives integer displacement of the soliton. It can be also applied to a strongly nonlinear regime, where the bands are intertwined with each other, leading to fractional displacement. That reminds us of the roles of interaction in the integer and fractional quantum Hall effects~\cite{Klitzing1980,Tsui1982}.
Our results reveal a new mechanism of how fictionalization emerges in the nonlinear-induced topological transitions and provide a new perspective on nonlinear topological materials~\cite{kirsch2021,Bai2024,Maczewsky2020}.
The correspondence between the nonlinear topological pumping and nonlinear bands can be verified in ultra-cold atom systems~\cite{Zhu2024} and photonic systems~\cite{zhang2022non}. 


\section{\label{sec:level1}Acknowledgments}

This work is supported by the National Natural Science Foundation of China (Grants No.~{12074367}, No.~{GG2030040453} and No.~{GG2030007011}), the Scientific Research Innovation Capability Support Project for Young Faculty (Gtant No.~{ZYGXQNJSKYCXNLZCXM-I26}), 
the Anhui Initiative in Quantum Information Technologies (Grant No.~{AHY040200}),
the National Key Research and Development Program of China (Grant No.~{2020YFA0309804}),
the Shanghai Municipal Science and Technology Major Project (Grant No.~{2019SHZDZX01}),
the Innovation Program for Quantum Science and Technology (Grants No.~{2021ZD0302002} and No.~{2021ZD0302004}),
the Strategic Priority Research Program of Chinese Academy of Sciences (Grant No.~{XDB35020200}),
and the New Cornerstone Science Foundation.



\begin{thebibliography}
\expandafter\ifx\csname url\endcsname\relax
\def\url#1{\texttt{#1}}\fi
\expandafter\ifx\csname urlprefix\endcsname\relax\def\urlprefix{URL }\fi
\providecommand{\bibinfo}[2]{#2}
\providecommand{\eprint}[2][]{\url{#2}}


\bibitem{Thouless1983}%
  \BibitemOpen
  \bibfield  {author} 
  {\bibinfo {author} {{D.~J.}~{Thouless}},}
  \href {https://doi.org/10.1103/PhysRevB.27.6083}
  {\bibfield{journal} {\bibinfo  {journal} {Phys. Rev. B}}
  \textbf{\bibinfo{volume} {27}}, \bibinfo {pages} {6083} (\bibinfo {year}{1983})}\BibitemShut {NoStop}%
\bibitem{Niu1984}%
  \BibitemOpen
  \bibfield  {author} {\bibinfo {author} { {Q.}~ {Niu}}
  and\bibinfo {author} { {D.~J.}~  {Thouless}},}
  \href {https://doi.org/10.1088/0305-4470/17/12/016} {\bibfield  {journal} {\bibinfo  {journal} {J. Phys. A: Math. Gen.} }
  \textbf {\bibinfo {volume} {17}},
  \bibinfo {pages} {2453}
  (\bibinfo {year} {1984})}\BibitemShut {NoStop}%
\bibitem{Niu1990}%
  \BibitemOpen
  \bibfield  {author} {\bibinfo {author} { {Q.}~ {Niu}},}
  \href {https://doi.org/10.1103/PhysRevLett.64.1812}
  {\bibfield  {journal} {\bibinfo  {journal} {Phys. Rev. Lett.}}
  \textbf {\bibinfo {volume} {64}},
  \bibinfo {pages} {1812}
  (\bibinfo {year} {1990})}\BibitemShut {NoStop}%
\bibitem{Citro2023}%
  \BibitemOpen
  \bibfield  {author}
  {\bibinfo {author} { {R.}~{Citro}}
  and \bibinfo {author} { {M.}~{Aidelsburger}},}
  \href {https://doi.org/10.1038/s42254-022-00545-0}
  {\bibfield  {journal} {\bibinfo  {journal} {Nat. Rev. Phys.}} 
  \textbf{\bibinfo {volume} {5}}, \bibinfo {pages} {87-101} (\bibinfo {year}{2023})}\BibitemShut {NoStop}%
\bibitem{Lohse2016}%
  \BibitemOpen
  \bibfield  {author}
  {\bibinfo {author} { {M.}~{Lohse}},
  \bibinfo {author} { {C.}~{Schweizer}},
  \bibinfo {author} { {O.}~{Zilberberg}}, 
  \bibinfo{author} { {M.}~{Aidelsburger}},
  and\bibinfo{author} { {I.}~{Bloch}},}
  \href{https://doi.org/10.1038/nphys3584} 
  {\bibfield  {journal} {\bibinfo{journal} {Nat. Phys.}}
  \textbf {\bibinfo {volume} {12}}, \bibinfo{pages} {350-354} (\bibinfo {year} {2016})} \BibitemShut {NoStop}%
\bibitem{Nakajima2016}%
  \BibitemOpen
  \bibfield  {author} 
  {\bibinfo {author} { {S.}~{Nakajima}},
  \bibinfo {author} { {T.}~{Tomita}},
  \bibinfo {author} { {S.}~{Taie}},
  \bibinfo {author}{ {T.}~{Ichinose}},
  \bibinfo {author}{ {H.}~{Ozawa}}, 
  \bibinfo {author} {{L.}~{Wang}}, 
  \bibinfo {author} { {M.}~{Troyer}},
  and \bibinfo {author} { {Y.}~{Takahashi}},}
  \href {https://doi.org/10.1038/nphys3622} 
  {\bibfield {journal} {\bibinfo  {journal} {Nat. Phys.}}
  \textbf {\bibinfo {volume}{12}}, \bibinfo {pages} {296-300} (\bibinfo {year} {2016})}\BibitemShut {NoStop}%
\bibitem{Lu2016}%
  \BibitemOpen
  \bibfield  {author} {\bibinfo {author} { {H.-I.}~{Lu}},
  \bibinfo {author} { {M.}~{Schemmer}},
  \bibinfo {author} { {L.~M.}~{Aycock}},
  \bibinfo{author} { {D.}~{Genkina}},
  \bibinfo {author}{ {S.}~{Sugawa}},
  and\bibinfo {author} { {I.~B.}~{Spielman}},}
  \href{https://doi.org/10.1103/PhysRevLett.116.200402} 
  {\bibfield  {journal}{\bibinfo  {journal} {Phys. Rev. Lett.}}
  \textbf {\bibinfo {volume}{116}}, \bibinfo {pages} {200402} (\bibinfo {year} {2016})}\BibitemShut {NoStop}%
\bibitem {Kraus2012}%
  \BibitemOpen
  \bibfield  {author} {\bibinfo {author} { {Y.~E.}~  {Kraus}},
  \bibinfo {author} { {Y.}~ {Lahini}},
  \bibinfo {author} { {Z.}~ {Ringel}},
  \bibinfo {author} { {M.}~ {Verbin}},
  and\ \bibinfo {author} { {O.}~ {Zilberberg}},}
  \href {https://doi.org/10.1103/PhysRevLett.109.106402}
  {\bibfield  {journal} {\bibinfo  {journal} {Phys. Rev. Lett.}}
  \textbf {\bibinfo {volume} {109}},~ \bibinfo {pages} {106402} (\bibinfo {year} {2012})} \BibitemShut {NoStop}%
\bibitem {Zilberberg2018}%
  \BibitemOpen
  \bibfield  {author} {\bibinfo {author} { {O.}~ {Zilberberg}},
  \bibinfo {author} { {S.}~ {Huang}},
  \bibinfo {author} { {J.}~ {Guglielmon}},
  \bibinfo {author} { {M.-H}~ {Wang}},
  \bibinfo {author} { {K.~P.}~  {Chen}},
  \bibinfo {author} { {Y.~E.}~  {Kraus}},
  and\ \bibinfo {author} { {M.~C.}~  {Rechtsman}},}
  \href {https://doi.org/10.1038/nature25011}
  {\bibfield  {journal} {\bibinfo  {journal} {Nature} }
  \textbf {\bibinfo {volume} {553}},
  \bibinfo {pages} {59-62} (\bibinfo {year} {2018})}\BibitemShut {NoStop}%
\bibitem{Cheng2020}%
  \BibitemOpen
  \bibfield  {author} {\bibinfo {author} { {W.-T}~ {Cheng}},
  \bibinfo {author} { {E.}~ {Prodan}},
  and\ \bibinfo {author} { {C.}~ {Prodan}},}
  \href {https://doi.org/10.1103/PhysRevLett.125.224301}
  {\bibfield  {journal} {\bibinfo  {journal} {Phys. Rev. Lett.}}
  \textbf {\bibinfo {volume} {125}},
  \bibinfo {pages} {224301}
  (\bibinfo {year} {2020})} \BibitemShut {NoStop}%
\bibitem{Fedorova2020}%
  \BibitemOpen
  \bibfield  {author} {\bibinfo {author} { {Z.}~ {Fedorova}},
  \bibinfo {author} {{H.-X}~ {Qiu}},
  \bibinfo {author} {{S.}~ {Linden}},
  and\bibinfo {author} { {J.}~ {Kroha}},}
  \href {https://doi.org/10.1038/s41467-020-17510-z}
  {\bibfield  {journal} {\bibinfo  {journal} {Nat. Commun.}}
  \textbf {\bibinfo {volume} {11}},~ \bibinfo {pages} {3758}
  (\bibinfo {year} {2020})}\BibitemShut {NoStop}%
\bibitem {Deng2024}%
  \BibitemOpen
  \bibfield  {author} {\bibinfo {author} { {C.-L.}~ {Deng}},
  \bibinfo {author} { {Y.}~ {Liu}},
  \bibinfo{author} { {Y.-R.}~  {Zhang}},
  \bibinfo {author}  { {X.-G.}~  {Li}},
  \bibinfo {author} {  {T.}~ {Liu}},
  \bibinfo {author} { {C.-T.}~   {Chen}},
   \bibinfo {author} { {T.}~  {Liu}},
  \bibinfo {author} { {C.-W.}~  {Lu}},
  \bibinfo {author} { {Y.-Y.}~  {Wang}},
  \bibinfo  {author} { {T.-M.}~  {Li}},
  \bibinfo {author}  { {C.-P.}~  {Fang}},
  \bibinfo {author}  { {S.-Y.}~  {Zhou}},
  \bibinfo {author}  { {J.-C.}~  {Song}},
  \bibinfo {author}  { {Y.-S.}~  {Xu}},
  \bibinfo {author} {  {Y.}~ {He}},
  \bibinfo {author} { {Z.-H.}~   {Liu}},
   \bibinfo {author} { {K.-X.}~  {Huang}},
  \bibinfo {author} { {Z.-C.}~  {Xiang}},
  \bibinfo {author} { {J.-C.}~ {Wang}},
  \bibinfo  {author} { {D.-N.}~  {Zheng}},
  \bibinfo {author}  { {G.-M.}~  {Xue}},
  \bibinfo {author} {  {K.}~ {Xu}},
  \bibinfo {author} { {H.~F.}~{Yu}},
  and\ \bibinfo {author} { {H.}~{Fan}},~}
  \href {https://link.aps.org/doi/10.1103/PhysRevLett.133.140402}
  {\bibfield  {journal} {\bibinfo  {journal} {Phys. Rev. Lett.}}
  \textbf {\bibinfo {volume} {133}},
  \bibinfo {pages} {140402}
  (\bibinfo {year} {2024})}  \BibitemShut {NoStop}%
\bibitem {Wu2019}%
  \BibitemOpen
  \bibfield  {author} {\bibinfo {author} { {Q.-S.} {Wu}}, \bibinfo {author} { {A.~A.}  {Soluyanov}},
  and\bibinfo {author} { {T.}~ {Bzdu{\v{s}}ek}},}
  \href {https://doi.org/10.1126/science.aau8740}
  {\bibfield  {journal}{\bibinfo  {journal} {Science}}
  \textbf {\bibinfo {volume} {365}},
  \bibinfo{pages} {1273-1277}
  (\bibinfo {year} {2019})} \BibitemShut {NoStop}%
\bibitem{You2022}%
  \BibitemOpen
  \bibfield  {author}
  {\bibinfo {author} { {O.-B.}~{You}},
  \bibinfo {author} { {S.-J.}~{Liang}},
  \bibinfo{author} { {B.-Y.}~{Xie}},
  \bibinfo {author}{ {W.-L.}~{Gao}},
  \bibinfo {author} {{W.-M.}~{Ye}},
  \bibinfo {author} { {J.}~{Zhu}},
  and\bibinfo {author} { {S.}~{Zhang}},}
  \href {https://doi.org/10.1103/PhysRevLett.128.244302}
  {\bibfield  {journal}{\bibinfo  {journal} {Phys. Rev. Lett.}}
  \textbf {\bibinfo {volume}{128}},
  \bibinfo {pages} {244302}
  (\bibinfo {year} {2022})}\BibitemShut {NoStop}%
\bibitem{Brosco2021}%
  \BibitemOpen
  \bibfield  {author}
  {\bibinfo {author} { {V.}~{Brosco}},
  \bibinfo {author} { {L.}~{Pilozzi}},
  \bibinfo {author} { {R.}~{Fazio}},
  and\bibinfo{author} { {C.}~{Conti}},}  \href{https://doi.org/10.1103/PhysRevA.103.063518}
  {\bibfield  {journal} {\bibinfo {journal} {Phys. Rev. A} }
  \textbf {\bibinfo {volume} {103}},
  \bibinfo{pages} {063518}
  (\bibinfo {year} {2021})}\BibitemShut {NoStop}%
\bibitem{Tangpanitanon2016}%
  \BibitemOpen
  \bibfield  {author} {\bibinfo {author} { {J.}~ {Tangpanitanon}},
  \bibinfo {author} { {V.~M.}~  {Bastidas}},
  \bibinfo {author} { {S.}~ {Al-Assam}},
  \bibinfo {author} { {P.}~ {Roushan}},
  \bibinfo {author} { {D.}~ {Jaksch}},
  and \bibinfo {author} { {D.~G.}~  {Angelakis}},}
  \href {https://doi.org/10.1103/PhysRevLett.117.213603}
  {\bibfield  {journal} {\bibinfo  {journal} {Phys. Rev. Lett.}}
  \textbf {\bibinfo {volume} {117}},
  \bibinfo {pages} {213603}
  (\bibinfo {year} {2016})}  \BibitemShut {NoStop}%
\bibitem{Leechaohong2020}%
  \BibitemOpen
  \bibfield  {author} {\bibinfo {author} { {L.}~{Lin}},
     \bibinfo {author} { {Y.-G.}~ {Ke}},
  and\bibinfo {author} { {C.-H.}~ {Lee}},}
  \href{https://doi.org/10.1103/PhysRevA.101.023620} 
  {\bibfield  {journal}{\bibinfo  {journal} {Phys. Rev. A.}}
  \textbf {\bibinfo {volume}{101}}, \bibinfo {pages} {023620} (\bibinfo {year} {2020})}\BibitemShut {NoStop}%
\bibitem{Leechaohong2023}%
  \BibitemOpen
  \bibfield  {author}
  {\bibinfo {author} { {Y.-G.}~ {Ke}},
  and\bibinfo {author} { {C.-H.}~ {Lee}},}
  \href {https://doi.org/10.1038/s41567-023-02169-2} 
  {\bibfield{journal} {\bibinfo  {journal} {Nat. Phys. }}
  \textbf {\bibinfo {volume}{19}},~ \bibinfo {pages} {1387-1388}
  (\bibinfo {year}{2023}{\natexlab{b}})}\BibitemShut {NoStop}%
\bibitem{Walter2023}%
  \BibitemOpen
  \bibfield  {author} {\bibinfo {author} { {A.~S.}~ {Walter}},
  \bibinfo {author} { {Z.-J.}~ {Zhu}},
  \bibinfo {author} { {M.}~ {G{\"{a}}chter}},
  \bibinfo {author} { {J.}~ {Minguzzi}},
  \bibinfo {author} { {S.}~ {Roschinski}},
  \bibinfo {author} { {K.}~ {Sandholzer}},
  \bibinfo {author} { {K.}~ {Viebahn}},
  and\bibinfo {author} { {T.}~ {Esslinger}},}
  \href {https://doi.org/10.1038/s41567-023-02145-w}
  {\bibfield  {journal} {\bibinfo  {journal} {Nat. Phys.}}
  \textbf {\bibinfo {volume} {19}},
  \bibinfo {pages} {1471-1475}
  (\bibinfo {year} {2023})}\BibitemShut {NoStop}%
\bibitem{Leechaohong2023b}%
  \BibitemOpen
  \bibfield  {author} {\bibinfo {author} { {Y.-G.}~ {Ke}},
      \bibinfo {author} { {J.-X.}~ {Huang}},
      \bibinfo {author} { {W.-J.}~ {Liu}},
      \bibinfo {author} { {Yuri}~ {Kivshar}},
  and\bibinfo {author} { {C.-H.}~ {Lee}},}
  \href{https://doi.org/10.1103/PhysRevLett.131.103604} 
  {\bibfield  {journal}{\bibinfo  {journal} {Phys. Rev. Lett.}}
  \textbf {\bibinfo {volume}{131}}, \bibinfo {pages} {103604} (\bibinfo {year} {2023})}\BibitemShut {NoStop}%
\bibitem{Viebahn2023}%
  \BibitemOpen
  \bibfield  {author} {\bibinfo {author} { {K.}~ {Viebahn}},
  \bibinfo {author} { {A.-S.}~  {Walter}},
  \bibinfo {author} { {E.}~ {Bertok}},
  \bibinfo {author} { {Z.-J.}~ {Zhu}},
  \bibinfo {author} { {M.}~ {G{\"{a}}chter}},
  \bibinfo {author} { {A.~A.}~  {Aligia}},
  \bibinfo {author} { {F.}~ {Heidrich-Meisner}},
  and\ \bibinfo {author} { {T.}~ {Esslinger}},}
  \href {https://link.aps.org/doi/10.1103/PhysRevX.14.021049}
  {\bibfield  {journal} {\bibinfo  {journal} {Phys. Rev. X}}
  \textbf {\bibinfo {volume} {14}},
  \bibinfo {pages} {021049}
  (\bibinfo {year} {2024})} \BibitemShut {NoStop}%
\bibitem {Liu2023}%
  \BibitemOpen
  \bibfield  {author} {\bibinfo {author} { {W.}~  {Liu}},
  \bibinfo {author} { {S.}~ {Hu}},
  \bibinfo  {author} { {L.}~ {Zhang}},
  \bibinfo {author}  { {Y.}~ {Ke}},
  and \bibinfo {author} { {C.}~ {Lee}},}
  \href {https://doi.org/10.1103/PhysRevResearch.5.013020}
  {\bibfield  {journal} {\bibinfo  {journal} {Phys. Rev. Res.}}
  \textbf {\bibinfo {volume} {5}}, \bibinfo {pages} {013020} (\bibinfo {year} {2023})}\BibitemShut {NoStop}%
\bibitem {Smirnova2020}%
  \BibitemOpen
  \bibfield  {author} {\bibinfo {author} { {D.}~  {Smirnova}},
  \bibinfo {author} { {D.}~ {Leykam}},
  \bibinfo {author} { {Y.}~ {Chong}},
  and \bibinfo {author} { {Y.}~ {Kivshar}},}
  \href {https://doi.org/10.1063/1.5142397}
  {\bibfield  {journal} {\bibinfo {journal} {Appl. Phys. Rev.}}
  \textbf {\bibinfo {volume} {7}}, \bibinfo {pages} {021306}
  (\bibinfo {year} {2020})}\BibitemShut {NoStop}%
\bibitem {Zhou2022}%
  \BibitemOpen
  \bibfield  {author} {\bibinfo {author} {{D.}~  {Zhou}},
  \bibinfo {author} {{D.~Z.}~  {Rocklin}},
  \bibinfo {author} {{M.}~ {Leamy}},
  and \bibinfo  {author} {{Y.}~ {Yao}}, }
  \href{https://doi.org/10.1038/s41467-022-31084-y} 
  {\bibfield  {journal} {\bibinfo  {journal} {Nat. Commun.} }
  \textbf {\bibinfo {volume}  {13}}, \bibinfo {pages} {3379} (\bibinfo {year} {2022})}\BibitemShut  {NoStop}%
\bibitem {Mukherjee2020}%
  \BibitemOpen
  \bibfield  {author} {\bibinfo {author} { {S.}~  {Mukherjee}}
  and \bibinfo {author} { {M.~C.}~  {Rechtsman}},}
  \href {https://doi.org/10.1126/science.aba8725}
  {\bibfield{journal} {\bibinfo  {journal} {Science}}
  \textbf {\bibinfo {volume}  {368}}, \bibinfo {pages} {856-859} (\bibinfo {year} {2020})} \BibitemShut {NoStop}%
\bibitem {Mukherjee2021}%
  \BibitemOpen
  \bibfield  {author} {\bibinfo {author} { {S.}~  {Mukherjee}}
  and \bibinfo {author} { {M.~C.}~ {Rechtsman}}, }
  \href {https://doi.org/10.1103/PhysRevX.11.041057}
  {\bibfield {journal} {\bibinfo  {journal} {Phys. Rev. X}}
  \textbf {\bibinfo {volume}  {11}}, \bibinfo {pages} {041057} (\bibinfo {year} {2021})}\BibitemShut {NoStop}%
\bibitem {Lumer2013}%
  \BibitemOpen
  \bibfield  {author} {\bibinfo {author} { {Y.}~  {Lumer}}, 
  \bibinfo {author} { {Y.}~ {Plotnik}},
  \bibinfo {author} { {M.~C.}~  {Rechtsman}},
  and  \bibinfo {author} { {M.}~ {Segev}}, }
  \href  {https://doi.org/10.1103/PhysRevLett.111.243905}
  {\bibfield  {journal}  {\bibinfo  {journal} {Phys. Rev. Lett.}}
  \textbf {\bibinfo {volume} {111}}, \bibinfo {pages} {243905} (\bibinfo {year} {2013})}\BibitemShut {NoStop}%
\bibitem {Leykam2016}%
  \BibitemOpen
  \bibfield  {author} {\bibinfo {author} { {D.}~  {Leykam}}
  and \bibinfo {author} { {Y.~D.}~ {Chong}}, }
  \href {https://doi.org/10.1103/PhysRevLett.117.143901}
  {\bibfield {journal} {\bibinfo  {journal} {Phys. Rev. Lett.}}
  \textbf {\bibinfo {volume} {117}}, \bibinfo {pages} {143901} (\bibinfo {year}  {2016})}\BibitemShut {NoStop}%
\bibitem {Marzuola2019}%
  \BibitemOpen
  \bibfield  {author} {\bibinfo {author} { {J.~L.}~ {Marzuola}}, \bibinfo {author} { {M.}~ {Rechtsman}},
  \bibinfo {author} { {B.}~ {Osting}},
  and \bibinfo {author} { {M.}~ {Bandres}},}
  \href {https://doi.org/10.48550/arXiv.1904.10312}
  {\bibfield  {journal}  {\bibinfo  {journal} {arXiv:1904.10312,}}
  (\bibinfo {year} {2019})}\BibitemShut {NoStop}%
\bibitem {Chaunsali2019}%
  \BibitemOpen
  \bibfield  {author} {\bibinfo {author} { {R.}~  {Chaunsali}}
  and \bibinfo {author} { {G.}~  {Theocharis}}, }
  \href {https://doi.org/10.1103/PhysRevB.100.014302}
  {\bibfield  {journal} {\bibinfo  {journal} {Phys. Rev. B} }
  \textbf {\bibinfo {volume} {100}}, \bibinfo {pages} {014302} 
  (\bibinfo {year}  {2019})}\BibitemShut {NoStop}%
\bibitem {Wong2023}%
  \BibitemOpen
  \bibfield  {author} {\bibinfo {author} {{S.}~  {Wong}},
  \bibinfo {author} {{T.~A.}~  {Loring}},
  and\ \bibinfo {author} {{A.}~ {Cerjan}}, }
  \href  {https://doi.org/10.1103/PhysRevB.108.195142} 
  {\bibfield  {journal} {\bibinfo {journal} {Phys. Rev. B} }
  \textbf {\bibinfo {volume} {108}}, \bibinfo  {pages} {195142} (\bibinfo {year} {2023})}\BibitemShut {NoStop}%
\bibitem{Jurgensen2021}%
  \BibitemOpen
  \bibfield  {author}
  {\bibinfo {author} { {M.}~{J{\"{u}}rgensen}},
  \bibinfo {author} { {S.}~{Mukherjee}},
  and\bibinfo {author} { {M.~C.}~{Rechtsman}},}
  \href {https://doi.org/10.1038/s41586-021-03688-9}
  {\bibfield{journal} {\bibinfo  {journal} {Nature}}
  \textbf {\bibinfo {volume}{596}},~ \bibinfo {pages} {63-67}
  (\bibinfo {year} {2021})}\BibitemShut{NoStop}%
\bibitem{Jurgensen2023}%
  \BibitemOpen
  \bibfield  {author}
  {\bibinfo {author} { {M.}~{J{\"{u}}rgensen}},
  \bibinfo {author} { {S.}~{Mukherjee}},
  \bibinfo {author} { {C.}~{J{\"{o}}rg}},
  and\bibinfo {author} { {M.~C.}~{Rechtsman}},}
  \href {https://doi.org/10.1038/s41567-022-01871-x} 
  {\bibfield{journal} {\bibinfo  {journal} {Nat. Phys. }}
  \textbf {\bibinfo {volume}{19}},~ \bibinfo {pages} {420-426}
  (\bibinfo {year}{2023}{\natexlab{b}})}\BibitemShut {NoStop}%
\bibitem{Fu2022}%
  \BibitemOpen
  \bibfield  {author}
  {\bibinfo {author} { {Q.-D.}~{Fu}},
  \bibinfo {author} { {P.}~ {Wang}},
  \bibinfo{author} { {Y.~V.}~{Kartashov}},
  \bibinfo {author}{ {V.~V.}~{Konotop}},
  and\bibinfo {author} { {F.-W.}~ {Ye}},}  \href{https://doi.org/10.1103/PhysRevLett.128.154101}
  {\bibfield  {journal}{\bibinfo  {journal} {Phys. Rev. Lett.}}
  \textbf {\bibinfo {volume}{128}},~ \bibinfo {pages} {154101}
  (\bibinfo {year}{2022}{\natexlab{a}})}\BibitemShut {NoStop}%
\bibitem{Jurgensen2022}%
  \BibitemOpen
  \bibfield  {author} 
  {\bibinfo {author} { {M.}~{J{\"{u}}rgensen}}
  and\bibinfo {author} { {M.~C.}~{Rechtsman}},}  \href{https://doi.org/10.1103/PhysRevLett.128.113901}
  {\bibfield  {journal}{\bibinfo  {journal} {Phys. Rev. Lett.} }
  \textbf {\bibinfo {volume}{128}},~ \bibinfo {pages} {113901}
  (\bibinfo {year} {2022})}\BibitemShut {NoStop}%
\bibitem{Tuloup2023}%
  \BibitemOpen
  \bibfield  {author}
  {\bibinfo {author} { {T.}~{Tuloup}}, 
  \bibinfo {author} { {R.~W.}~{Bomantara}},
  and\bibinfo {author} { {J.}~{Gong}},}
  \href{https://doi.org/10.1088/1367-2630/acef4d}
  {\bibfield  {journal} {\bibinfo  {journal} {New J. Phys.}}
  \textbf {\bibinfo {volume} {25}},
  \bibinfo{pages} {083048}
  (\bibinfo {year} {2023})}\BibitemShut {NoStop}%
\bibitem {Mostaan2022}%
  \BibitemOpen
  \bibfield  {author} {\bibinfo {author} { {N.}~ {Mostaan}},
  \bibinfo {author} { {F.}~ {Grusdt}},
  and\bibinfo {author} { {N.}~ {Goldman}},}
  \href {https://doi.org/10.1038/s41467-022-33478-4}
  {\bibfield  {journal} {\bibinfo  {journal} {Nat. Commun.}}
  \textbf {\bibinfo {volume} {13}},~ \bibinfo {pages} {5997}
  (\bibinfo {year} {2022})} \BibitemShut {NoStop}%
\bibitem {Yuan2023}%
  \BibitemOpen
  \bibfield  {author} {\bibinfo {author} { {T.}~{Yuan}},
  \bibinfo {author} { {H.-N.}~ {Dai}},
  and\  \bibinfo {author} { {Y.-A.}~ {Chen}}, }
  \href {https://doi.org/10.7498/aps.72.20230740}
  {\bibfield  {journal} {\bibinfo  {journal} {Acta Phys. Sin.}}
  \textbf {\bibinfo {volume} {72(16)}}:~\bibinfo {pages} {160302}, (\bibinfo {year} {2023})}\BibitemShut {NoStop}%
\bibitem{Fu2022a}%
  \BibitemOpen
  \bibfield  {author}
  {\bibinfo {author} { {Q.-D.}~{Fu}}, 
  \bibinfo {author} { {P.}~{Wang}},
  \bibinfo{author} { {Y.~V.}~{Kartashov}},
  \bibinfo {author}{ {V.~V.}~{Konotop}},
  and\bibinfo {author}{ {F.-W.}~ {Ye}},}  \href{https://doi.org/10.1103/PhysRevLett.129.183901}
  {\bibfield  {journal}{\bibinfo  {journal} {Phys. Rev. Lett.}}
  \textbf {\bibinfo {volume}{129}},~ \bibinfo {pages} {183901}
  (\bibinfo {year}{2022}{\natexlab{b}})}\BibitemShut {NoStop}%
\bibitem{Leechaohong2025}%
  \BibitemOpen
  \bibfield  {author} {\bibinfo {author} { {X.-R.}~{You}},
      \bibinfo {author} { {L.-Y.}~ {Xiao}},
      \bibinfo {author} { {B.-N.}~ {Huang}},
      \bibinfo {author} { {Y.-G.}~ {Ke}},
  and\bibinfo {author} { {C.-H.}~ {Lee}},}
  \href{https://doi.org/10.1103/PhysRevA.111.033306} 
  {\bibfield  {journal}{\bibinfo  {journal} {Phys. Rev. A.}}
  \textbf {\bibinfo {volume}{111}}, \bibinfo {pages} {033306} (\bibinfo {year} {2025})}\BibitemShut {NoStop}%
\bibitem {Zhang2009}%
  \BibitemOpen
  \bibfield  {author} {\bibinfo {author} { {Y.}~{Zhang}},
  \bibinfo {author} { {Z.}~ {Liang}},
  and  \bibinfo {author} { {B.}~ {Wu}},}
  \href{https://doi.org/10.1103/PhysRevA.80.063815} 
  {\bibfield  {journal} {\bibinfo{journal} {Phys. Rev. A}}
  \textbf {\bibinfo {volume} {80}}, \bibinfo{pages} {063815} (\bibinfo {year} {2009})}\BibitemShut {NoStop}%
\bibitem {Cao2024}%
  \BibitemOpen
  \bibfield  {author} {\bibinfo {author} { {X.}~  {Cao}},
  \bibinfo {author} { {C.}~ {Jia}},
  \bibinfo {author} { {Y.}~ {Hu}},
  and \bibinfo {author}{ {Z.}~ {Liang}},}
  \href  {https://doi.org/10.1103/PhysRevA.110.013305}
  {\bibfield  {journal} {\bibinfo{journal} {Phys. Rev. A} }
  \textbf {\bibinfo {volume} {110}}, \bibinfo{pages} {013305} (\bibinfo {year} {2024})}\BibitemShut {NoStop}%
\bibitem {Ravets2025}%
  \BibitemOpen
  \bibfield  {author} {\bibinfo {author} { {S.}~  {Ravets}},
  \bibinfo {author} { {N.}~ {Pernet}},
  \bibinfo {author} { {N.}~ {Mostaan}},
  \bibinfo {author} { {N.}~ {Goldman}},
  and \bibinfo {author} { {J.}~ {Bloch}}, }
  \href  {https://doi.org/10.1103/PhysRevLett.134.093801}
  {\bibfield  {journal}{\bibinfo  {journal} {Phys. Rev. Lett.}}
  \textbf {\bibinfo {volume} {134}},  \bibinfo {pages} {093801} (\bibinfo {year} {2025})}\BibitemShut {NoStop}%
\bibitem {Tao2025}%
  \BibitemOpen
  \bibfield  {author} {\bibinfo {author} { {Y.-L.}~  {Tao}}, \bibinfo {author} { {Y.}~ {Zhang}},
  and \bibinfo {author} { {Y.}~ {Xu}}, }
  \href  {https://doi.org/10.48550/arXiv.2502.06131}
  {{arXiv:2502.06131} }
  (\bibinfo {year} {2025})\BibitemShut {NoStop}%
\bibitem {Cao2025}%
  \BibitemOpen
  \bibfield  {author} {\bibinfo {author} { {X.}~  {Cao}},
  \bibinfo {author} { {C.}~ {Jia}},
  \bibinfo {author} { {H.}~ {Lyu}}, 
  \bibinfo {author} { {Y.}~ {Hu}},
  and \bibinfo {author} { {Z.}~ {Liang}}, }
  \href {https://doi.org/10.1103/PhysRevA.111.023329}
  {\bibfield  {journal} {\bibinfo {journal} {Phys. Rev. A}}
  \textbf {\bibinfo {volume} {111}}, \bibinfo  {pages} {023329} (\bibinfo {year} {2025})}\BibitemShut {NoStop}%
\bibitem{Sone2024}%
  \BibitemOpen
  \bibfield  {author} {\bibinfo {author} { {K.}~
  {Sone}},
  \bibinfo {author} { {M.}~ {Ezawa}},
  \bibinfo {author} { {Y.}~ {Ashida}},
  \bibinfo {author} { {N.}~ {Yoshioka}},
  and\bibinfo {author} { {T.}~ {Sagawa}},}
  \href{https://doi.org/10.1038/s41567-024-02451-x}
  {\bibfield  {journal}{\bibinfo  {journal} {Nat. Phys.}
  \textbf {\bibinfo {volume} {20}},
  \bibinfo{pages} {1164–1170}
  (\bibinfo {year} {2024})}} \BibitemShut {NoStop}%
\bibitem{Wu2000}%
  \BibitemOpen
  \bibfield  {author} {\bibinfo {author} { {B.}~ {Wu}}
  and \bibinfo {author} { {Q.}~ {Niu}},}
  \href {https://doi.org/10.1103/PhysRevA.61.023402} {\bibfield  {journal} {\bibinfo  {journal} {Phys. Rev. A} }
  \textbf {\bibinfo {volume} {61}},
  \bibinfo {pages} {023402}
  (\bibinfo {year} {2000})} \BibitemShut {NoStop}%
\bibitem{Supplementary}
    \BibitemOpen
    {See Supplemental Material for the modified Newton method, the degenerate bands of the three-site AAH model with strong nonlinearity, the Thouless pumping of solitons in the seven-site nonlinear AAH model and the corresponding nonlinear Chern number, the energy structures of the ill-defined regions in five- and seven- site nonlinear AAH models, and the 3D phase diagram for the Rice-Mele model. }
\bibitem{Wilczek1984}%
  \BibitemOpen
  \bibfield  {author} {\bibinfo {author} { {F.}~ {Wilczek}}
  and\bibinfo {author} { {A.}~ {Zee}},}
  \href {https://doi.org/10.1103/PhysRevLett.52.2111} {\bibfield  {journal} {\bibinfo  {journal} {Phys. Rev. Lett.}}
  \textbf {\bibinfo {volume} {52}},
  \bibinfo {pages} {2111}
  (\bibinfo {year} {1984})}\BibitemShut {NoStop}%
\bibitem{Fukui2005}%
  \BibitemOpen
  \bibfield  {author}
  {\bibinfo {author} { {T.}~{Fukui}}, 
  \bibinfo {author} { {Y.}~ {Hatsugai}},
  and \bibinfo {author} { {H.}~ {Suzuki}},}
  \href{https://doi.org/10.1143/jpsj.74.1674} 
  {\bibfield  {journal} {\bibinfo{journal} {J. Phys. Soc. Jpn.}}
  \textbf {\bibinfo{volume} {74}},
  \bibinfo {pages} {1674-1677}
  (\bibinfo {year}{2005})}\BibitemShut {NoStop}%
\bibitem{Thouless1982}%
  \BibitemOpen
  \bibfield  {author} {\bibinfo {author} { {D.~J.}~  {Thouless}},
  \bibinfo {author} { {M.}~ {Kohmoto}},
  \bibinfo {author} { {M.~P.}~  {Nightingale}},
  and\ \bibinfo {author} { {M.}~ {den Nijs}},}
  \href {https://doi.org/10.1103/PhysRevLett.49.405} {\bibfield  {journal} {\bibinfo  {journal} {Phys. Rev. Lett.}}
  \textbf {\bibinfo {volume} {49}},
  \bibinfo {pages} {405}
  (\bibinfo {year} {1982})} \BibitemShut {NoStop}%
\bibitem {Klitzing1980}%
  \BibitemOpen
  \bibfield  {author} {\bibinfo {author} { {K.~v.}~{Klitzing}},
  \bibinfo {author} { {G.}~ {Dorda}},
  and\ \bibinfo {author} { {M.}~ {Pepper}},~}
  \href{https://doi.org/10.1103/PhysRevLett.45.494} {\bibfield  {journal} {\bibinfo
  {journal} {Phys. Rev. Lett.}}~\textbf {\bibinfo {volume} {45}},~\bibinfo
  {pages} {494} (\bibinfo {year} {1980})}\BibitemShut {NoStop}%
\bibitem {Tsui1982}%
  \BibitemOpen
  \bibfield  {author} {\bibinfo {author} { {D.~C.}~{Tsui}},
  \bibinfo {author} { {H.~L.}~{Stormer}},
  and\ \bibinfo {author} { {A.~C.}~{Gossard}},~}
  \href {https://doi.org/10.1103/PhysRevLett.48.1559} {\bibfield  {journal}
  {\bibinfo  {journal} {Phys. Rev. Lett.}~}\textbf {\bibinfo {volume} {48}},
  \bibinfo {pages} {1559} (\bibinfo {year} {1982})}\BibitemShut {NoStop}%
\bibitem {Maczewsky2020}%
  \BibitemOpen
  \bibfield  {author} {\bibinfo {author} { {L.~J.}~{Maczewsky}},
  \bibinfo {author} { {M.}~ {Heinrich}},
  \bibinfo {author} { {M.}~ {Kremer}},
  \bibinfo{author} { {S.~K.}~{Ivanov}},
  \bibinfo {author}{ {M.}~ {Ehrhardt}},
  \bibinfo {author}{ {F.}~ {Martinez}},
  \bibinfo {author}{ {Y.~V.}~{Kartashov}},
  \bibinfo {author}{ {V.~V.}~{Konotop}},
  \bibinfo {author}{ {L.}~ {Torner}},
  \bibinfo {author} {{D.}~ {Bauer}},
  and\ \bibinfo {author} {{A.}~ {Szameit}},~}
  \href{https://doi.org/10.1126/science.abd2033} {\bibfield  {journal} {\bibinfo
  {journal} {Science}~}\textbf {\bibinfo {volume} {370}},
  \bibinfo {pages}{701-704} (\bibinfo {year} {2020})}\BibitemShut{NoStop}%
\bibitem {Bai2024}%
  \BibitemOpen
  \bibfield  {author} {\bibinfo {author} {{K.}~  {Bai}},
  \bibinfo {author} {{J.-Z.}~  {Li}},
  \bibinfo {author} {{T.-R.}~  {Liu}},
  \bibinfo {author} {{L.}~ {Fang}},
  \bibinfo {author} {{D.}~ {Wan}},
  and \bibinfo {author}  {{M.}~ {Xiao}},}
  \href  {https://doi.org/10.1103/PhysRevLett.133.116602}
  {\bibfield  {journal}  {\bibinfo  {journal} {Phys. Rev. Lett.}}
  \textbf {\bibinfo {volume} {133}},  \bibinfo {pages} {116602} (\bibinfo {year} {2024})}\BibitemShut {NoStop}%
\bibitem{kirsch2021}%
  \BibitemOpen
  \bibfield  {author} {\bibinfo {author} { {M.~S.}~ {Kirsch}},
  \bibinfo {author} { {Y.-Q.}~ {Zhang}},
  \bibinfo {author} { {M.}~ {Kremer}},
  \bibinfo{author} { {L.~J.}~  {Maczewsky}},
  \bibinfo {author}{ {S.~K.}~  {Ivanov}}, 
  \bibinfo {author}{ {Y.~V.}~  {Kartashov}}, 
  \bibinfo {author}{ {L.}~ {Torner}},
  \bibinfo {author} {{D.}~ {Bauer}}, 
  \bibinfo {author} {{A.}~ {Szameit}},
  and\ \bibinfo {author} {{M.}~ {Heinrich}},~ }
  \href {https://doi.org/10.1038/s41567-021-01275-3} {\bibfield  {journal}
  {\bibinfo  {journal} {Nat. Phys.}~ }\textbf {\bibinfo {volume} {17}},
  \bibinfo {pages} {995-1000} (\bibinfo {year} {2021})}\BibitemShut {NoStop}%
\bibitem{Zhu2024}%
    \BibitemOpen
    \bibfield  {author} {\bibinfo {author} { {Z.-J.}~ {Zhu}},
    \bibinfo {author} { {Y.}~  {Kiefer}},
    \bibinfo {author} { {S.}~  {Jele}},
    \bibinfo {author} { {M.}~  {G\"{a}chter}},
    \bibinfo {author} { {G.}~  {Bisson}},
    \bibinfo {author} { {K.}~  {Viebahn}},
   and\ \bibinfo {author} { {T.}~  {Esslinger}},~}
      \href {https://arxiv.org/pdf/2409.02984}
      {{arXiv:2409.02984}
      (\bibinfo {year} {2024})}\BibitemShut {NoStop}%
\bibitem {zhang2022non}%
  \BibitemOpen
  \bibfield  {author} {\bibinfo {author} { {X.-L.}~ {Zhang}},
  \bibinfo {author} { {F.}~ {Yu}},
  \bibinfo {author} { {Z.-G.}~  {Chen}},
  \bibinfo {author}{ {Z.-N.}~  {Tian}},
  \bibinfo {author}{ {Q.-D.}~  {Chen}},
  \bibinfo {author}{ {H.-B.}~  {Sun}},
  and\ \bibinfo {author}{ {G.-C}~ {Ma}},~}
  \href {https://doi.org/10.1038/s41566-022-00976-2} {\bibfield  {journal}{\bibinfo  {journal} {Nat. Photon.}~ }\textbf {\bibinfo {volume} {16}},
  \bibinfo {pages} {390-395} (\bibinfo {year} {2022})}\BibitemShut {NoStop}\end{thebibliography}

\bibliographystyle{apsrev4-2}


%



\setcounter{equation}{0}
\setcounter{figure}{0}
\setcounter{table}{0}
\setcounter{page}{1}
\setcounter{section}{0}
\makeatletter
\makeatother
\global\def\theequation{S\arabic{equation}}
\global\def\thefigure{S\arabic{figure}}
\global\def\thetable{S\arabic{table}}
\global\def\thepage{S\arabic{page}}
\global\def\thesection{S\arabic{section}}

\clearpage

\setcounter{equation}{0}
\setcounter{figure}{0}
\setcounter{table}{0}
\setcounter{page}{1}
\setcounter{section}{0}
\makeatletter
\makeatother
\global\def\theequation{S\arabic{equation}}
\global\def\thefigure{S\arabic{figure}}
\global\def\thetable{S\arabic{table}}
\global\def\thepage{S\arabic{page}}
\global\def\thesection{S\arabic{section}}
\renewcommand{\bibnumfmt}[1]{[#1]}
\renewcommand{\citenumfont}[1]{#1}

\onecolumngrid
\begin{center}
    {\Large Supplementary material for\\
    \vspace{0.2em}
    ``{\bf \thetitle}''
    }\\
    \vspace{0.5em}
    Fei-Fei~{Wu {\it et. al.}}\\
     \vspace{0.2em}
    (\today)\\
    \vspace{1em}
\end{center}

\onecolumngrid

\section{\label{sec:level1} Modified Newton Method}

Due to the nonlinear eigenvectors, the results may be larger than the dimensions of the Hamiltonian.
To get all the instantaneous eigenvectors and eigenvalues, a numerical approach, Newton's down-hill method is employed to avoid non-converging situations of random initial inputs.
Here's a brief outline of the iterative procedure:

1. Start with a random vector, $\Phi_{\mathrm{guess}}$, as the initial guess for the eigenvector;

2. Use $\Phi_{\mathrm{guess}}$ in the Newton's down-hill method.
Check whether the results converge once the number of iterations hits the maximum limit and only the converged results are retained;

3. Repeat steps 1 and 2 multiple times to obtain a set of eigenvectors and eigenvalues for the instant Hamiltonian.

The number of variables is linearly dependent on the size of the unit cell, leading to the search of all eigenvectors difficult and large computation for the lager model.
To overcome this problem, we first try plenty of times with random initial guesses for several special $\{k,t\}$ points. Then we collect all the results as the preknowledge and use them as the initial guesses to calculate the eigenstates. We update the initial guesses with the eigenstates of $\{k,t\}$, and use them to the neighbor points. To make sure that we do not miss any other states beyond the initial guesses, we maintain about 100 random guesses for every $\{k,t\}$ and perform a post-examination of the data.

\begin{figure}[hbtp]
	\includegraphics{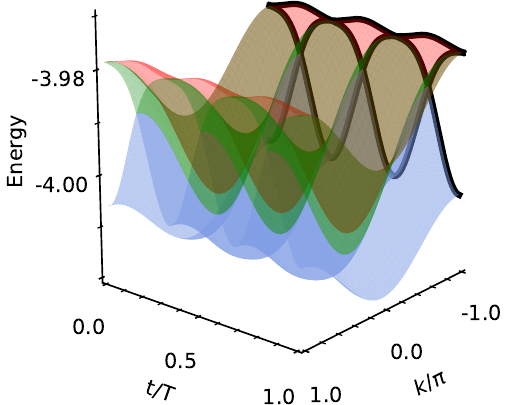}
	\caption{\label{FigA1} 
    The ground states for the three-site nonlinear AAH model of $g\!=\!4.0$.
    Three degenerate states are showed by blue, green and red, respectively.
    Black lines are energy bands for $k=-\pi$.
    }
\end{figure}
\begin{figure}[hbtp]
	\includegraphics{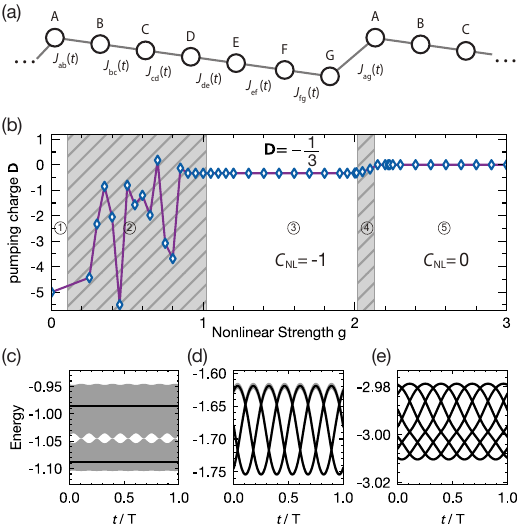}
	\caption{\label{FigA2} 
    (a), Schematic of the unit cell with seven sites.
    (b), The phase diagram and pumped charge for the nonlinear seven-site AAH model.
    There are three topological phases noted by circled 1, 3, and 5.
    The gray regions 2 and 4 are the ill-defined areas.
    The blue diamonds are the pumped charges.
    (c), Some low energy bands correspond to the three well-defined regions in (b).
    From left to right is of $g\! =\! 0.0$, $g\! =\! 1.5$ and $g\! =\!  3.0$.
    For $g\! =\! 1.5$, the lowest three bands are shown. For $g\! =\! 3.0$, the lowest seven bands are braiding.
    The gray shadows are the projection of bands over $k$.
    The black lines are the energy bands at $k\!=\!0.5\pi$.
    The tunneling parameters are $J\! =\! 1.2$ and $K\! =\! 1.0$ for all figures above. 
    }
\end{figure}
\section{\label{sec:level1} The degenerate bands for the strong nonlinearity}
In the manuscript, we show nonlinear bands on a large scale for the strongly nonlinear region(Fig.2(g)). 
Here, the ground states are plotted on a small scale to demonstrate the degeneracy.
In Fig~\ref{FigA1}, they consists of three degenerate states with degenerate points at $t=\{0,~\pi/3,~2\pi/3,~\pi,~4\pi/3,~5\pi/3\}$.
For example, at $t=0$, the upper two bands are degenerate for all momentum parameter $k$; at $t=\pi$, the lowest two bands are degenerate for all $k$.

\section{\label{sec:level1} Seven-site nonlinear AAH model}
Further extending the size of the unit cell to seven, the $-1/3$ pumped charge occurs.
In Fig~\ref{FigA2}, we show the displacement of the dynamical pumping simulation and the band structures of the lowest several bands.
The tunneling parameter is modulated by  $J_{j,j+1}(t) \!=\! \left[J+K \cos \left(6 \pi j/7+\Omega t-3 \pi/14\right)\right]/(J+K)$, where $J = 1.2$ and $K = 1.0$. 
There are three topological phases: one has an isolated nonlinear ground state ( Fig~\ref{FigA2}(c) ) while the other two have braiding ground states( Fig~\ref{FigA2}(d-e) ).
In region \textcircled{1} where $C_{\mathrm{NL}}\!=\!-5$, the solitons are not formed;
in region \textcircled{3} where $C_{\mathrm{NL}}\!=\!-1$, the ground states are three braiding bands and the pumped charge is $\mathbf{D} = C_{\mathrm{NL}}/3 \!=\! -1/3$, which agrees well with dynamical simulations;
in region \textcircled{5}, the Chern number is $C_{\mathrm{NL}}=0$ with $N$=5 degenerate space, and the solitons are localized, which agree well with Eq. (1) in the manuscript, $\mathbf{D} = C_{\mathrm{NL}}/5 = 0$.

\begin{figure}[htbp]
	\includegraphics{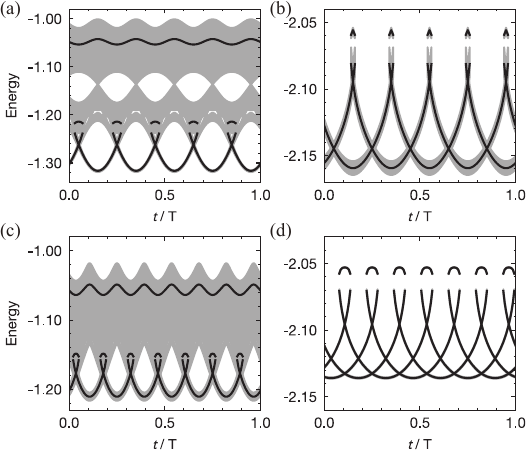}
	\caption{\label{FigA3} 
	(a) and (b), The lowest several band structures of $g=0.6$ and $g=2.1$ for the five-site AAH model.
    (c) and (d), The lowest several band structures of $g=0.4$ and $g=2.1$ for the seven-site AAH model.
}
\end{figure}
\begin{figure}[htbp]
	\includegraphics{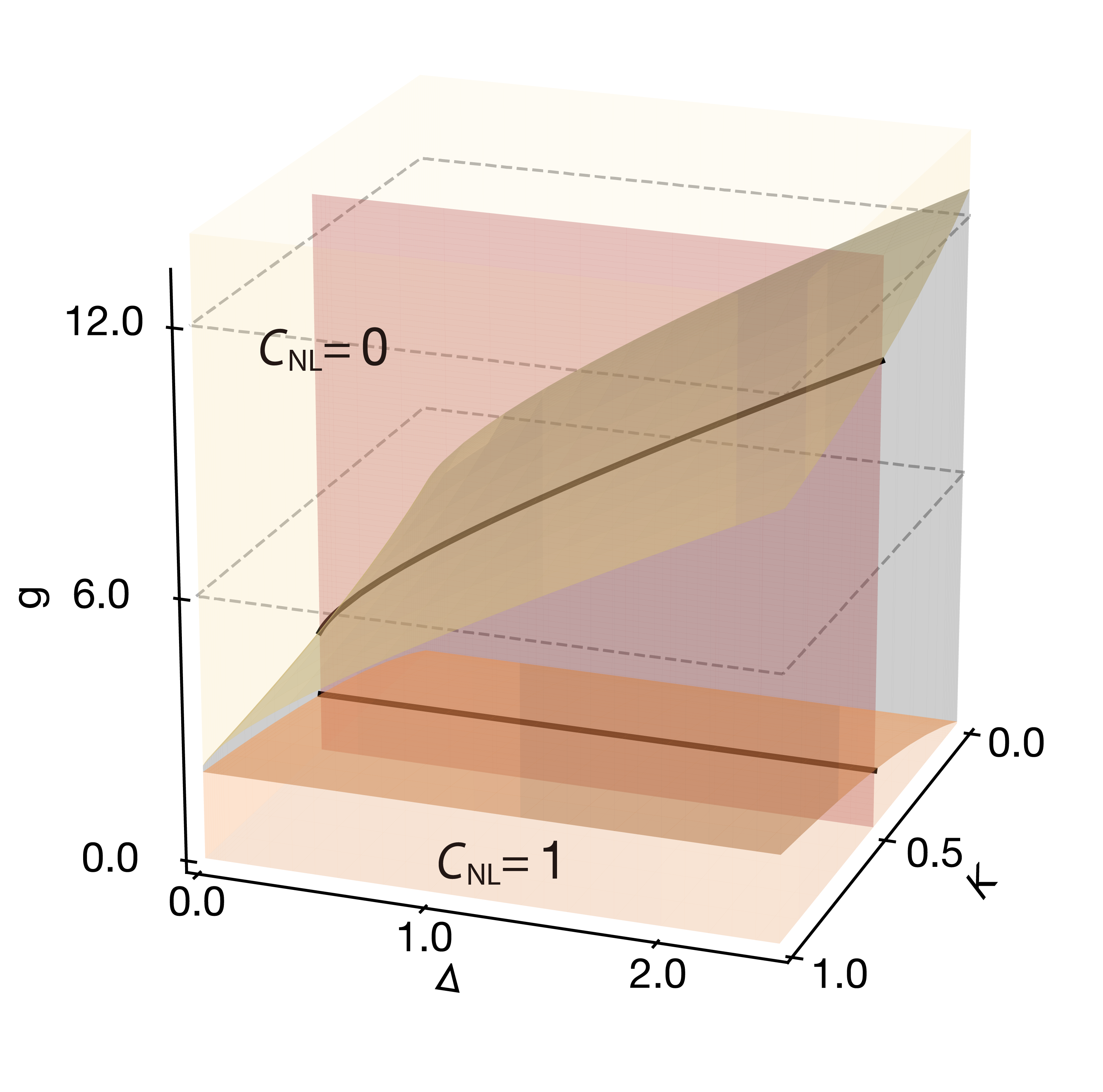}
	\caption{\label{FigA4} 
    The phase diagram of the nonlinear RM model.
    The black lines are the phase boundaries for $K=0.5$.
    }
\end{figure}

\section{\label{sec:level1} Energy Structure of the ill-Defined Region}
In this section, we show the lowest several bands for the regions \textcircled{2} and \textcircled{4} in Fig.3(b) of the manuscript and Fig.~\ref{FigA2}(b).
These bands in ill-defined regions are degenerate but not completely filled 2D BZ.
As shown in Fig~\ref{FigA3}(a), it is of $g=0.6$ for the five-site AAH model.
The lowest band is degenerate with a loop band, which occurs only in some areas of BZ.
In Fig~\ref{FigA3}(b), the nonlinear strength increases to $g=2.1$.
The two entirely filled degenerate bands touch a loop band.

Fig~\ref{FigA3}(c) and Fig~\ref{FigA3}(d) are of $g=0.4$ and $g=2.1$ for the seven-site AAH model.
We find more complicated band structures but also obey the touch of non-filled bands and the lowest bands.
The black lines are the energy bands for $k=0.5\pi$.

\section{\label{sec:level1} Phase diagram of the nonlinear Rice-Mele model}
Here, we show the three-dimensional topological phase diagram of the nonlinear Rice-Mele model.
The horizontal axes are the modulation amplitudes $\Delta$ and $K$ for linear onsite energy and the tunneling strength.
The vertical axis is the nonlinear strength $g$.
The whole parameter space is divided into three parts (filled by different colors in Fig~\ref{FigA4}), two of which are well-defined regions and the other is an ill-defined region .
With the increasing of the nonlinear strength, the topological invariant transfer from $C_{\mathrm{NL}}\!=\! 1$ to $C_{\mathrm{NL}} \!=\! 0$.
The phase boundaries are shown by two curved surfaces in Fig~\ref{FigA4}, which are related to both the linear and nonlinear parameters.

\end{document}